\documentclass{article} 
\usepackage[super,compress]{cite}
\usepackage{amsmath}
\usepackage{amssymb}
\usepackage{graphicx}

\title{Momentum Distribution and Correlation due to mass difference caused by power-like distribution}
\author{Masamichi Ishihara}

\def\<#1>{\langle #1 \rangle}
\def\vevapp#1{\langle #1 \rangle^{\mathrm{MFPA}}}
\def\ketvec#1{\left| #1\right\rangle}
\def\bravec#1{\left\langle #1\right|}

\begin{document}
\maketitle
\begin{abstract}
The momentum distribution and particle correlation due to the mass difference
were studied both in the case of the conventional expectation value and in the case of $q$-expectation value, 
when the momentum distribution is described by a Tsallis distribution with the entropic parameter $q \ge 1$. 
The magnitude of the momentum distribution for hard modes increases as $q$ increases, 
and the $q$-dependence of the momentum distribution is quite weak for soft modes.
The correlation at $q>1$ is larger than that at $q=1$ for soft modes, 
while the correlation at $q>1$ is smaller than that at $q=1$ for hard modes. 
The $q$-dependence of these quantities in the case of $q$-expectation value is weaker 
than that in the case of the conventional expectation value, respectively.
\end{abstract}



\section{Introduction}
\label{sec:introduction}

Power-like distributions appear in various branches of science, and have been studied by many researchers. 
An probable power-like distribution is a Tsallis distribution which has two parameters:
the temperature $T$ and the entropic parameter $q$.
This distribution is applied to many phenomena, such as particle distributions in high energy heavy ion collisions.  
It was shown that the momentum distribution in high energy collisions is described well 
by the Tsallis distribution with $q>1$ \cite{wilk2007,osada2008,cleymans2012,marques2015}. 

An extended statistics of the Boltzmann-Gibbs (BG) statistics is the Tsallis statistics \cite{tsallis-book}.
The definition of the expectation value in the Tsallis statistics \cite{tsallis1998,lavagno2002} differs that in the BG statistics. 
Therefore, the expectation value in the Tsallis statistics differs that in the BG statistics even when the distribution is the same,  
because the expectation value depends on the definition in the statistics.
The difference of the statistics is significant when the physical values are calculated. 

The effective mass is affected by the distribution and the statistics.
Therefore, the effective mass was calculated in the Tsallis statistics \cite{Rozynek2009,Santos2014,ishihara2016}.
The mass affected by the Tsallis distribution is different from that by the BG distribution. 
The effective mass in the Tsallis statistics \cite{ishihara2016} is also different from that in the BG statistics \cite{ishihara2015} 
even when the distribution is the same, 
because  the definition of the expectation value in the Tsallis statistics is  different from that in the BG statistics.
The phenomena caused by mass is affected by the distribution and the statistics.

The mass difference causes the particle production and the correlation \cite{birrell:book}.
It was shown that a particle with momentum $\vec{k}$ is correlated with a particle with momentum $-\vec{k}$ \cite{asakawa1999, padula2006}.
Therefore, the momentum distribution and the correlation between particles are affected by the distribution and the statistics. 

The purpose of this paper is to clarify the momentum distribution and the correlation due to mass difference 
caused by the difference of the distribution and that by the difference of the statistics.
The mass is affected by the distribution and the statistics in high energy collisions.
In this paper, pion mass is calculated in the linear sigma model.
The momentum distribution and the correlation are studied, because these quantities are affected by the mass difference.

This paper is organized as follows. 
In section~\ref{sec:mom-corr}, 
we derive the momentum distribution and the correlation due to mass difference. 
The effective pion mass is also estimated when the distribution is power-like.
In section~\ref{sec:numerical-estimation},
the momentum distribution and the correlation are numerically estimated at high energies.
Last section is assigned for discussion and conclusion. 

\section{Momentum distribution and correlation}
\label{sec:mom-corr}
\subsection{Momentum distribution and  correlation caused by the mass difference}
In the present calculation, we assume that the mass $m$ changes at time $t=0$:
\begin{align}
m = \left\{ \begin{array}{ll} m_{+} & \qquad (t > 0) \\ m_{-} & \qquad (t<0) \end{array} \right. , 
\end{align}
where the positive sign indicates $t>0$ and the negative sign for $t<0$.

We introduce the energy $\omega_{\pm}(\vec{k}) = \sqrt{\vec{k}^2 + m_{\pm}^2}$ and  
annihilation operator $a_{\pm} (\vec{k})$ with the commutation relations,
$\left[a_{\pm}(\vec{k}) ,  a_{\pm}^{\dag}(\vec{l}) \right]=\delta(\vec{k}-\vec{l})$ and $\left[a_{\pm}(\vec{k}) ,  a_{\pm}(\vec{l}) \right]= 0$.
The field $\phi(x)$ is expanded:
\begin{equation}
\phi(x) = 
\displaystyle\int \ 
\frac{d\vec{k}}{\sqrt{(2\pi)^3 (2 \omega_{\pm}(\vec{k}))}} \left. \left( a_{\pm}(\vec{k}) e^{-i k x} + a_{\pm}^{\dag}(\vec{k}) e^{i k x}\right) \right|_{k^0 = \omega_{\pm}(\vec{k})}  , 
\end{equation}
where $kx = k^0 t - \vec{k} \cdot \vec{x}$.

The requirements that the field $\phi(x)$ and the derivative $d\phi(x)/dt$ are continuous at $t=0$, lead to the following relations. 
\begin{subequations}
\begin{align}
a_{+}(\vec{k}) & = \cosh(\theta(\vec{k})) a_{-}(\vec{k}) + \sinh(\theta(\vec{k})) a^{\dag}_{-}(-\vec{k}) ,\\
a_{+}^{\dag}(\vec{k}) & = \cosh(\theta(\vec{k})) a_{-}^{\dag}(\vec{k}) + \sinh(\theta(\vec{k})) a_{-}(-\vec{k}) , 
\end{align}
\label{eqn:a+a-relation}
\end{subequations}
where $\cosh(\theta(\vec{k}))$ and $\sinh(\theta(\vec{k}))$ are given by 
\begin{subequations}
\begin{align}
\cosh(\theta(\vec{k})) &= \frac{1}{2} \left( \sqrt{\frac{\omega_{+}(\vec{k})}{\omega_{-}(\vec{k})}}  + \sqrt{\frac{\omega_{-}(\vec{k})} {\omega_{+}(\vec{k})}} \right) ,\\
\sinh(\theta(\vec{k})) &= \frac{1}{2} \left( \sqrt{\frac{\omega_{+}(\vec{k})}{\omega_{-}(\vec{k})}}  - \sqrt{\frac{\omega_{-}(\vec{k})} {\omega_{+}(\vec{k})}} \right) .
\end{align}
\label{eqn:angle-relation}
\end{subequations}
The ground state $\ketvec{0,\pm}$ is defined by $a_{\pm}(\vec{k}) \ketvec{0,\pm} = 0$.

The momentum distribution is calculated when the mass changes at $t=0$.
We assume that the momentum distribution is a Tsallis distribution $f^{(r)}(\vec{k}; T, q)$ at $t<0$: 
\begin{subequations}
\begin{align}
\< a_{-}^{\dag}(\vec{k})  a_{-}(\vec{l}) >^{(r)} = \left[ f^{(r)}(\vec{k}; T, q) \right]^r \delta(\vec{k} - \vec{l})  
\qquad (r=1, q).
\label{mom:def}
\end{align}
where $\< {\cal O} >$ represents the statistical average of the quantity ${\cal O}$ for the equilibrium at $t<0$.
 This average is simply given by $\bravec{0,-} {\cal O} \ketvec{0,-}$ when the temperature is $0$ at $t<0$. 
The superscript denoted as $(r)$ distinguishes between the conventional expectation value with a Tsallis distribution and the $q$-expectation value. 
The value in the case of the conventional expectation value with a Tsallis distribution corresponds to $r=1$, 
and the value in the case of the $q$-expectation value to $r=q$.
The energies $\omega_{\pm}$ depend on the mass which is $r$-dependent, and we attach the superscript $(r)$  to $\omega_{\pm}$.
Therefore, the angle $\theta$ is also $r$-dependent: $\theta  = \theta^{(r)}$.
The following expectation value for free particle satisfies 
\begin{align}
\< a_{-}(\vec{k})  a_{-}(\vec{l}) >^{(r)} = 0 . 
\end{align}
\end{subequations}

The Tsallis distribution function $f^{(r)}(\vec{k} ; T, q)$ for boson is given by 
\begin{equation}
f^{(r)}(\vec{k}; T, q) = \frac{1}{\left[ 1 + (q-1) \left( \frac{\omega^{(r)}_{-}(\vec{k})}{T} \right) \right]_{+}^{\frac{1}{(q-1)}} -1 }
,
\end{equation}
where $k := | \vec{k} |$.  
The notation $[x]_{+}$ represents $x$ for $x\ge 0$ and $0$ for $x<0$.
The parameter $T$ is called temperature in this paper, 
because it is just the temperature of Boltzmann-Gibbs statistics when $q$ is equal to 1. 
The parameter $q$ is called entropic parameter.
It is evident that the upper limit of $k$ exists in the case of $q<1$.  

Hereafter, we focus only on the  case of $q>1$, 
because it is reported in high energy collisions that the value of $q$ is larger than or equal to $1$.

The quantity $\<  a_{+}^{\dag}(\vec{k}) a_{+}(\vec{l}) >^{(r)}$ is given by 
\begin{align}
& \<  a_{+}^{\dag}(\vec{k}) a_{+}(\vec{l}) >^{(r)}  
\nonumber \\ &
= \left[ \left( \cosh^2(\theta^{(r)}(\vec{k})) + \sinh^2(\theta^{(r)}(\vec{k}) ) \right) \left[ f^{(r)}(\vec{k}; T, q) \right]^r 
+  \sinh^2(\theta^{(r)}(\vec{k}))  \right] \delta(\vec{k}-\vec{l}) , 
\end{align}
and the momentum distribution in the unit volume is 
\begin{equation}
\frac{1}{V} \frac{dN}{dk} 
= \frac{k^2}{2\pi^2}  \left[ \left( \cosh^2(\theta^{(r)}(\vec{k})) + \sinh^2(\theta^{(r)}(\vec{k}) ) \right) \left[ f^{(r)}(\vec{k}; T, q) \right]^r 
+  \sinh^2(\theta^{(r)}(\vec{k}))  \right] 
.
\label{eqn:momdist}
\end{equation}


The following correlation $C^{(r)}(\vec{k},\vec{l})$ is calculated:
\begin{align}
C^{(r)}(\vec{k},\vec{l}) 
= \frac{\<a_{+}^{\dag}(\vec{k})  a_{+}^{\dag}(\vec{l}) a_{+}(\vec{k}) a_{+}(\vec{l})>^{(r)}}{\<  a_{+}^{\dag}(\vec{k}) a_{+}(\vec{k}) >^{(r)} \< a_{+}^{\dag}(\vec{l}) a_{+}(\vec{l}) >^{(r)}}
.
\end{align}
We apply the following approximation to the operators constructed from $a_{-}^{\dag}(\vec{k})$ and $a_{-}(\vec{k})$
for free particle when $q$ is close to 1, referring to the generalized Wick's theorem \cite{sinyukov1994,sinyukov1999}.
For example, 
\begin{align}
\< a_{-}(\vec{k}) a^{\dag}_{-}(\vec{l}) a_{-}(\vec{p}) a_{-}^{\dag}(\vec{q})>^{(r)} 
& \sim  
\< a_{-}(\vec{k}) a^{\dag}_{-}(\vec{l})>^{(r)} \< a_{-}(\vec{p}) a_{-}^{\dag}(\vec{q})>^{(r)}
\nonumber \\  & \quad 
+\< a_{-}(\vec{k}) a_{-}(\vec{p}) >^{(r)} \< a^{\dag}_{-}(\vec{l}) a_{-}^{\dag}(\vec{q})>^{(r)}
\nonumber \\  & \quad 
+ \< a_{-}(\vec{k}) a_{-}^{\dag}(\vec{q}) >^{(r)} \< a^{\dag}_{-}(\vec{l}) a_{-}(\vec{p}) >^{(r)} . 
\end{align}
The correlations under the above approximation are given by 
\begin{subequations}
\begin{align}
&C^{(r)}(\vec{k},\vec{l})_{\vec{k} \neq \vec{l}} = 1 , \\
&C^{(r)}(\vec{k},\vec{k})_{\vec{k} \neq \vec{0}} = 2 ,\\
&C^{(r)}(\vec{k},-\vec{k})_{\vec{k} \neq \vec{0} } 
\nonumber \\ &
= 1 + \frac{\cosh^2 (\theta^{(r)}(\vec{k})) \sinh^2 (\theta^{(r)}(\vec{k}))   \left( 2 \left[ f^{(r)}(\vec{k}; T, q) \right]^r  + 1 \right)^2}
{\left\{ \cosh^2 (\theta^{(r)}(\vec{k})) \left[ f^{(r)}(\vec{k}; T, q) \right]^r + \sinh^2 (\theta^{(r)}(\vec{k}))  \left( \left[ f^{(r)}(\vec{k}; T, q) \right]^r + 1 \right) \right\}^2}
, \label{eqn:k-k} \\
&C^{(r)}(\vec{0},\vec{0}) 
\nonumber \\ &
= 2 + \frac{\cosh^2 (\theta^{(r)}(\vec{0})) \sinh^2 (\theta^{(r)}(\vec{0}))   \left( 2 \left[ f^{(r)}(\vec{0}; T, q) \right]^r + 1 \right)^2}
{\left\{ \cosh^2 (\theta^{(r)}(\vec{0})) \left[ f^{(r)}(\vec{0}; T, q) \right]^r + \sinh^2 (\theta^{(r)}(\vec{0}))  \left( \left[ f^{(r)}(\vec{0}; T, q) \right]^r + 1 \right)\right\}^2}
.
\end{align}
\end{subequations}

We focus on the distribution, Eq.~\eqref{eqn:momdist}, and the correlation, Eq.~\eqref{eqn:k-k}, in the next section.

\subsection{Effective masses in the linear sigma model}
The linear sigma model is described with $N$ scalar fields, $\phi_0$, $\phi_1$, $\cdots$, $\phi_{N-1}$. 
The Hamiltonian density of the linear sigma model is 
\begin{equation}
{\cal H} = \frac{1}{2} \left( \partial^0 \phi \right)^2 + \frac{1}{2} \left( \nabla \phi \right)^2 
                   + \frac{\lambda}{4} \left(  \phi^2 - v^2 \right)^2  - G \phi_0  ,
\end{equation}
where $\phi^2 = \sum_{i=0}^{N-1} \phi_i^2$, 
$\left( \partial^0 \phi \right)^2 =  \sum_{i=0}^{N-1}  \left( \partial^0 \phi_i \right)^2$, 
and $\left( \nabla \phi \right)^2 =  \sum_{i=0}^{N-1}  \left( \nabla \phi_i \right)^2$. 
The parameter $\lambda$ is a coupling constant and $- G \phi_0$ represents the symmetry breaking term.

The Hamiltonian density after an expectation value is taken is calculated:  
The field $\phi_i$ is divided as $\phi_i = \phi_{ic} + \phi_{ih}$, 
where $\phi _{ic}$ is the condensate of the field $\phi_i$ and $\phi_{ih}$ is the remaining part. 
We take the expectation value with respect to $\phi_{ih}$ under the massless free particle approximation.

The expectation value $\<\left( \phi_{ih} \right)^2>$ is estimated to obtain masses.
A Tsallis distribution in the massless case is given by 
\begin{equation}
\left. f^{(r)}(\vec{k}; T, q) \right|_{\omega^{(r)}_{-} = k} =  \frac{1}{\left[ 1 + (q-1) \left( \frac{k}{T} \right) \right]_{+}^{\frac{1}{(q-1)}} -1 } .
\end{equation}
The value $\<\phi_{ih}>$ is independent of suffix $i$ under the massless free particle approximation (MFPA) \cite{gavin1994,ishihara1999}.   
We introduce $K_q^{(r)}$ to simplify the equation:
\begin{equation}
K^{(r)} (T, q) := \vevapp{\left( \phi_{ih} \right)^2} = \int \frac{d\vec{k}}{(2\pi)^3} \frac{1}{k} \left[ \left. f^{(r)}(\vec{k}; T, q) \right|_{\omega^{(r)}_{-}=k} \right]^r  
\qquad  (r=1, q), 
\end{equation}
The value in the case of the conventional expectation value with a Tsallis distribution is given by using $K^{(1)}(T,q)$ 
and the value in the case of the q-expectation value is given by using $K^{(q)}(T,q)$.

The potential is tilted to the $\phi_0$ direction.
The field $\phi_{i\mathrm{c}}$ ($i \neq 0$) is zero, and $\phi_{0\mathrm{c}}$ satisfies the following equation:
\begin{equation}
\left( \phi_{\mathrm{0c}} \right)^3 + \left[(N+2) K^{(r)} (T,q) - v^2 \right] \phi_{\mathrm{0c}} - G/\lambda = 0 
\qquad  (r = 1, q).
\end{equation}
We represent the solution of the above equation as $\phi_{\mathrm{0c}}^{(r)}$. 
The mass squared $\left(m_i^{(r)}\right)^2$ is given by \cite{ishihara2016,ishihara2015} 
\begin{equation}
\left( m_i^{(r)} \right) ^2 = \lambda \left[ ( 1 + 2 \delta_{i0}) \left( \phi_{\mathrm{0c}}^{(r)} \right)^2 + (N+2) K^{(r)} (T,q) - v^2 \right]   
\qquad  (r = 1, q). 
\end{equation}
The mass $m_i^{(r)}$ is used to calculate the momentum distribution and correlation.

\section{Numerical estimation}
\label{sec:numerical-estimation}
We apply the method in the case of  mass modification at high energy heavy ion collisions.
We set the parameters of the model:
the number of fields $N$ is set to 4, 
and the parameters $\lambda$, $v$, and $G^{1/3}$ are set to 20, 89.4 MeV, and 119 MeV respectively.
At zero temperature, 
these values of the parameters correspond to $m_0^{(r)}=600$ MeV, $m_i^{(r)} = 135$ MeV ($i =1,2,3$), and pion decay constant $f_{\pi} = 92.5$ MeV. 

The quantity $m_{i}$ $(i=1,2,3)$ is pion mass. 
In the present situation, the pion mass for $t<0$ is $m_{i}^{(r)}(T,q)$ and the mass for $t>0$ is $m_{i}^{(r)}(T=0,q)$. 
That is,
\begin{eqnarray}
m^{(r)}_i = 
\left\{
\begin{array}{ll} 
m_{i}^{(r)}(T=0,q) & \qquad (t>0)\\ 
m_{i}^{(r)}(T,q) & \qquad (t<0)          
\end{array}
\right.
.
\end{eqnarray}

\begin{figure}
\begin{center}
\includegraphics[width=0.45\textwidth]{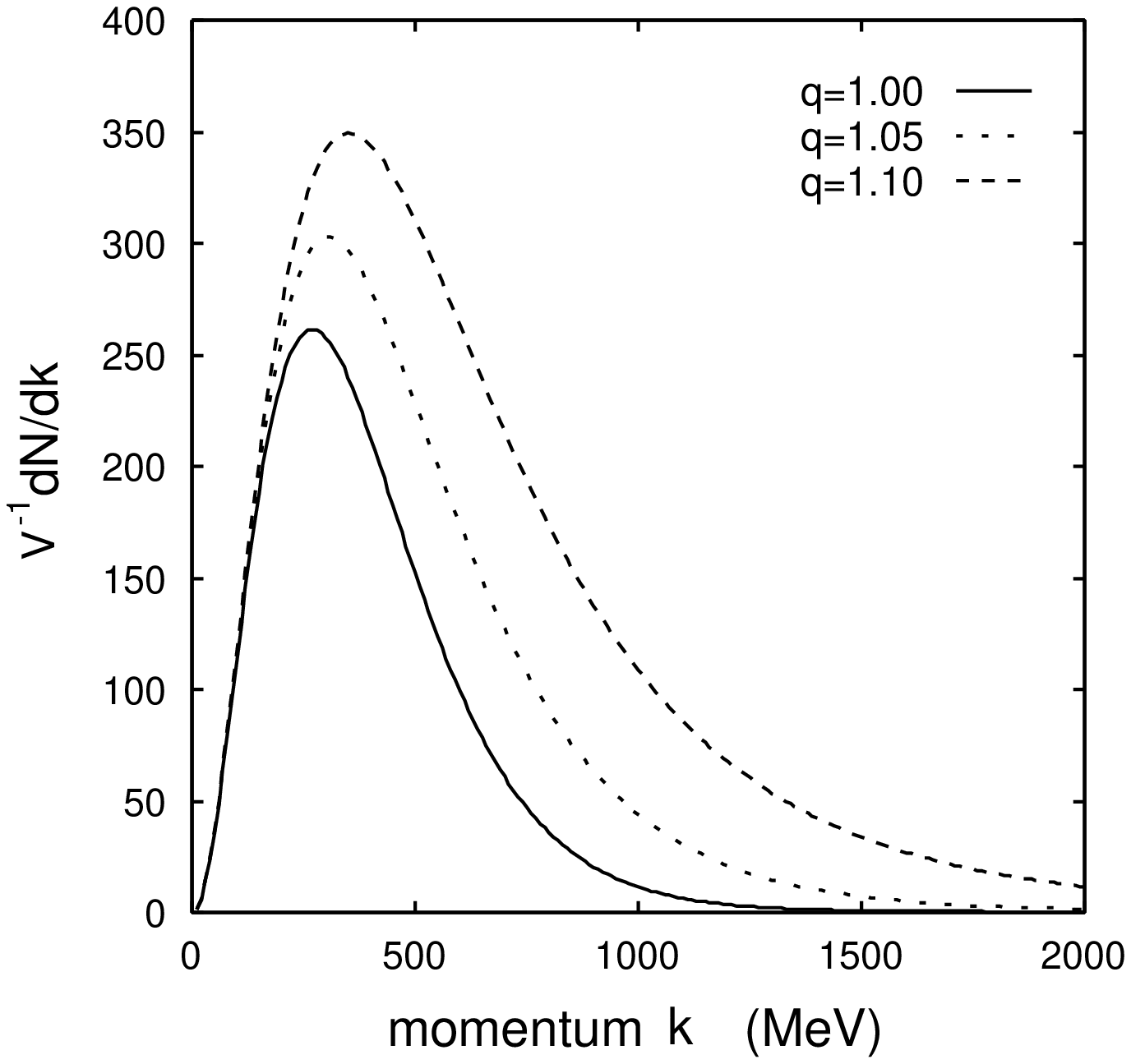}
\includegraphics[width=0.45\textwidth]{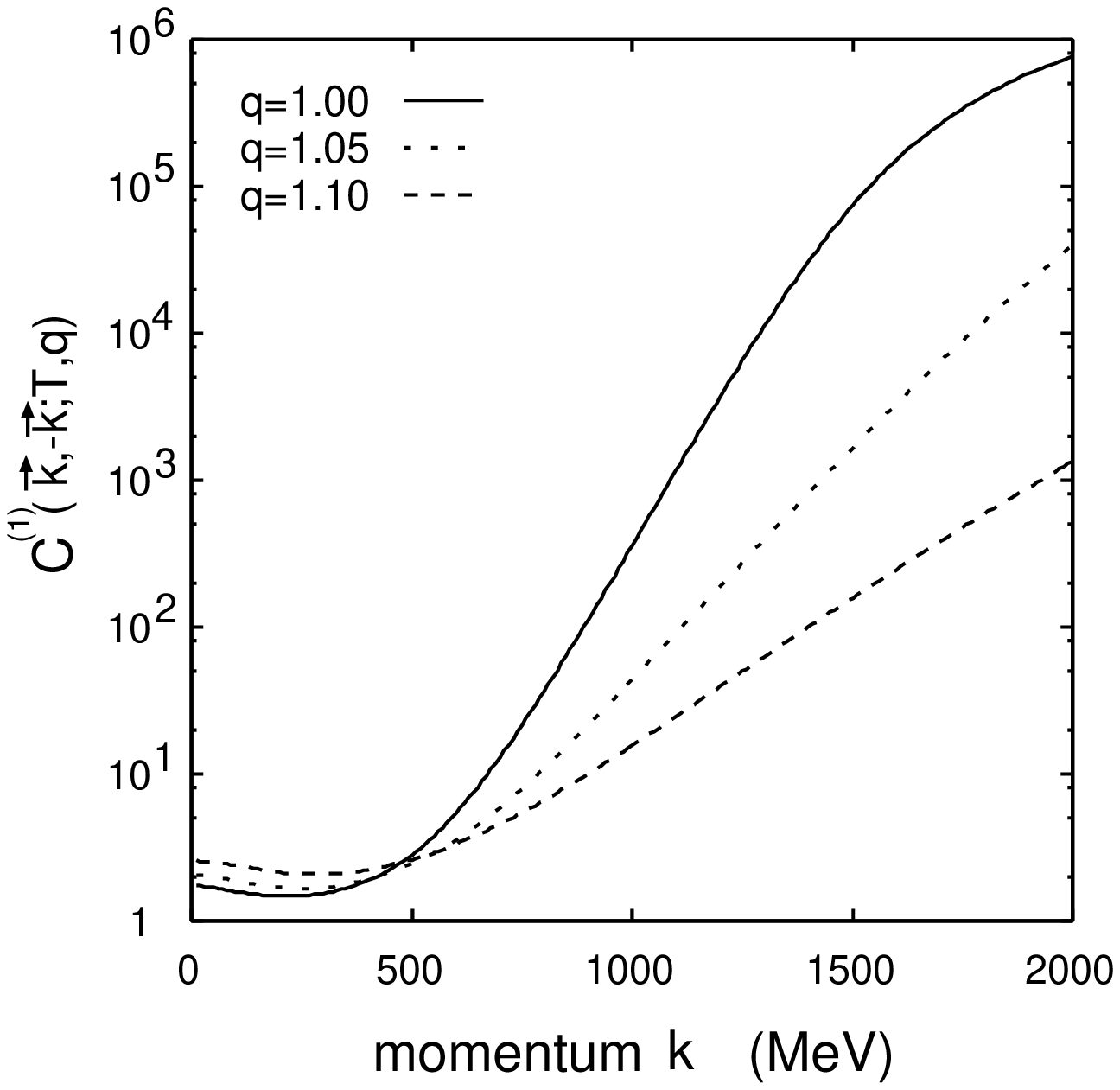}
\end{center}
\caption{
Momentum distribution in the unit volume $V^{-1} dN/dk$ (left panel) and correlation $C^{(1)}(\vec{k}, -\vec{k};T, q)$ (right panel)
in the range of $0$ MeV $< k \le$ $2000$ MeV at $T=120$ MeV for $q=1.00, 1.05$, and $1.10$ in the case of the conventional expectation value.
}
\label{Fig:conventional}
\end{figure}

First, the momentum distribution in the unit volume $V^{-1} dN/dk$ 
and correlation $C^{(r)}(\vec{k}, -\vec{k}; T, q)$ for pion were calculated for various $T$ and $q$ in the case of conventional expectation value ($r=1$), 
where we describe the parameters $T$ and $q$ explicitly. 
Figure~\ref{Fig:conventional} shows the quantities for pion at $T=120$ MeV.
The entropic parameter $q$ is $1.00$, $1.05$, and $1.10$. 
The shapes of these curves are quite similar.  
We represent the momentum at the peak of the distribution as $k_p$.
The value $k_p$ is generally $q$-dependent.
In Fig.~\ref{Fig:conventional}, the value is approximately $500$ MeV.
The magnitude of the momentum distribution becomes to be large for $k>k_p$ as $q$ becomes to be large, 
while the magnitude does not change for $k<k_p$. 
The variation of the correlation softens as $q$ increases. 
The value of the correlation at $q>1$ is larger than that at $q=1$ for small $k$, 
and that at $q>1$ is smaller than that at $q=1$ for large $k$. 
The correlation,  eq.~\eqref{eqn:k-k}, diverges at $k=\infty$,  because $f^{(r)}(\vec{k}; T, q)$ is zero and $\sinh \theta^{(r)}(\vec{k})$ is zero at $k = \infty$.

\begin{figure}
\begin{center}
\includegraphics[width=0.45\textwidth]{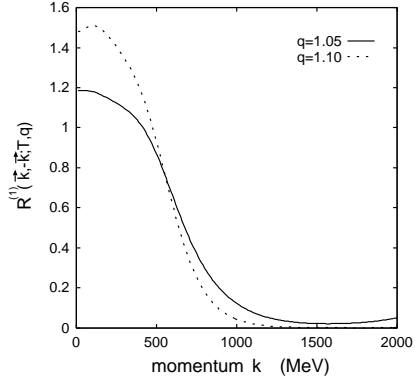}
\end{center}
\caption{The ratio $R^{(1)}(\vec{k}, -\vec{k}; T, q)$ in the range of $0$ MeV$< k \le$ $2000$ MeV 
at $T=120$ MeV for $q=1.05$ and $q=1.10$ in the conventional expectation value.}
\label{Fig:dist:ratio-corr}
\end{figure}

Figure~\ref{Fig:dist:ratio-corr} shows the ratio $R^{(1)}(\vec{k}, -\vec{k};T, q)$ 
at $T=120 \mathrm{MeV}$ for $q=1.05$ and $1.10$ in the conventional expectation value, where the ratio is defined by 
\begin{align}
R^{(r)}(\vec{k}, -\vec{k}; T, q) := \frac{C^{(r)}(\vec{k}, -\vec{k}; T, q)}{C^{(r)}(\vec{k}, -\vec{k}; T, q=1) }  \qquad (r=1, q ) .
\end{align}
The ratio at $q>1$ is smaller than $1$ for large $k$.
This fact is easily explained by the mass difference. 
The mass at $q>1$ is heavier than the mass at $q=1$ for $t<0$: $m_i^{(r)}(t,q>1)  >  m_i^{(r)}(t,q=1)$.
Therefore, $|\sinh \theta^{(r)}(\vec{k})|$ at $q>1$ is larger than $|\sinh \theta^{(r)}(\vec{k})|$ at $q=1$ from eq.~\eqref{eqn:angle-relation}.
This indicates from eq.~\eqref{eqn:k-k}  that the correlation at $q>1$ is smaller than that at $q=1$ for large $k$, 
because the distribution function can be ignored for large $k$. 

\begin{figure}
\begin{center}
\includegraphics[width=0.45\textwidth]{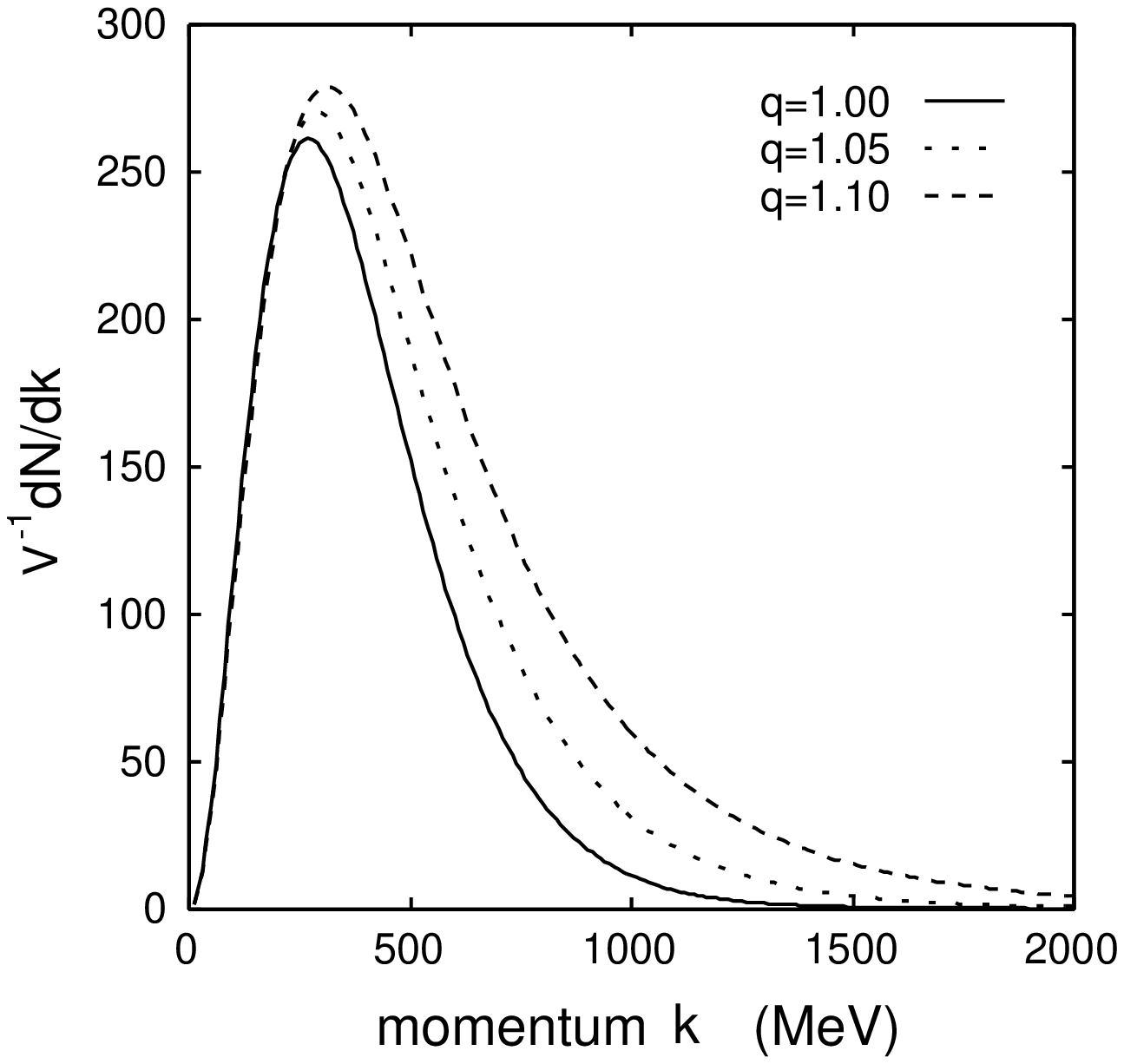}
\includegraphics[width=0.45\textwidth]{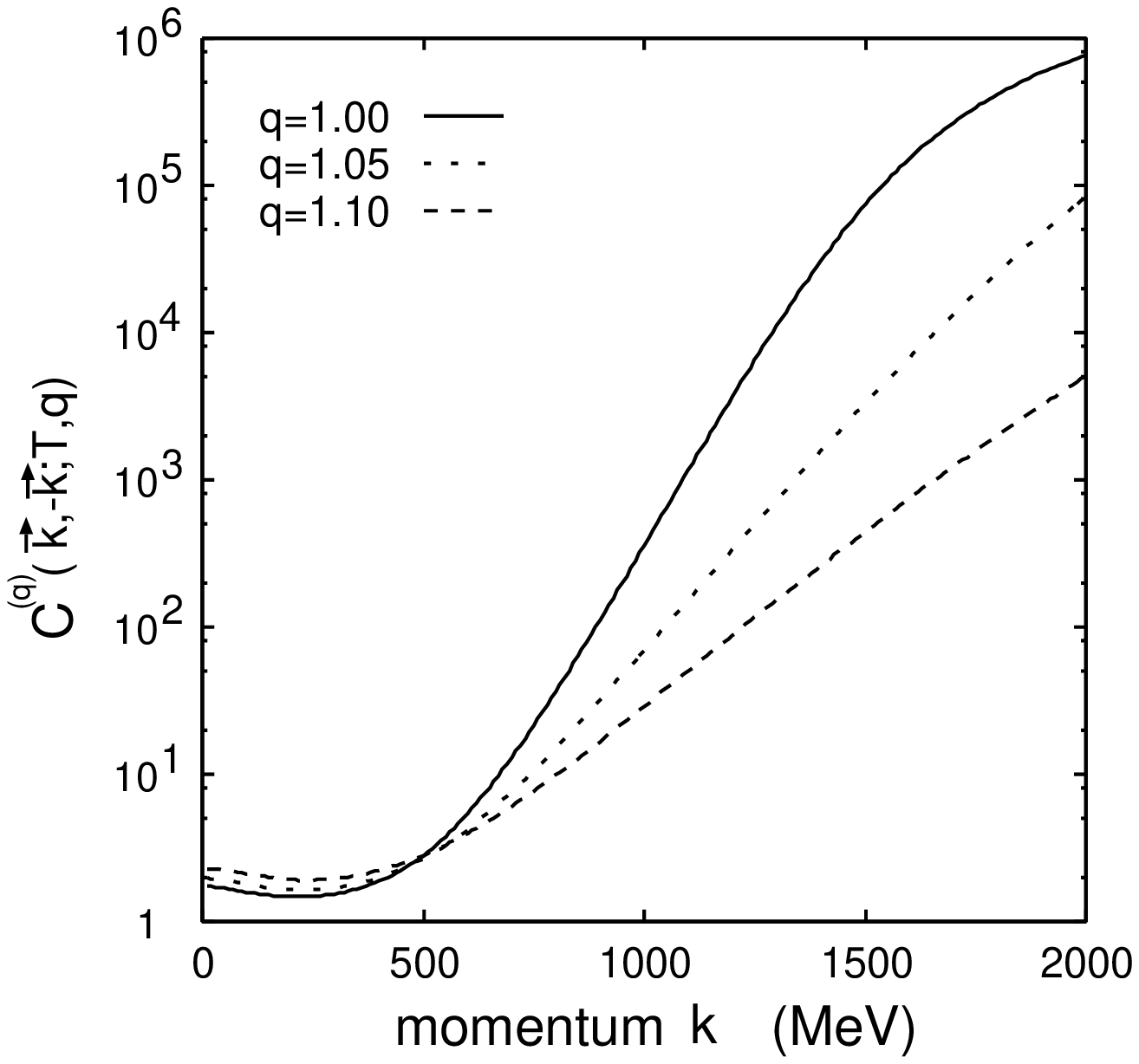}
\end{center}
\caption{
Momentum distribution in the unit volume $V^{-1} dN/dk$ (left panel) and correlation $C^{(q)}(\vec{k}, -\vec{k}; T, q)$ (right panel) 
in the range of $0$ MeV$< k \le$ $2000$ MeV at $T=120$ MeV for $q=1.00, 1.05$, and $1.10$ in the case of the $q$-expectation value.
}
\label{Fig:q-expectation}
\end{figure}

Second, 
these quantities were calculated  for various $T$ and $q$ in the case of $q$-expectation value ($r=q$). 
Figure~\ref{Fig:q-expectation} shows the quantities for pion at $T=120$ MeV.
The entropic parameter $q$ is $1.00$, $1.05$, and $1.10$. 
The behavior in the $q$-expectation value is similar to that in the conventional expectation value. 
The $q$-dependence of the momentum distribution in the $q$-expectation value 
is weaker than that in the conventional expectation value.  
Similarly, 
the $q$-dependence of the correlation in the $q$-expectation value 
is weaker than that in the conventional expectation value.  

\begin{figure}
\begin{center}
\includegraphics[width=0.45\textwidth]{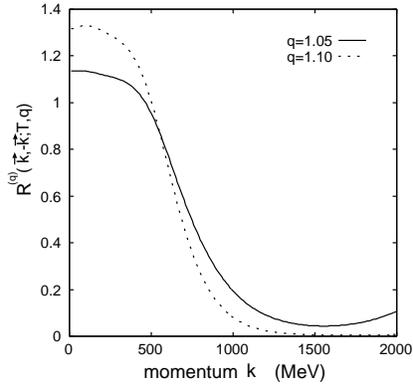}
\end{center}
\caption{The ratio $R^{(q)}(\vec{k}, -\vec{k}; T, q)$ in the range of $0$ MeV$< k \le$ $2000$ MeV
at $T=120$ MeV for $q=1.05$ and $q=1.10$ in the $q$-expectation value.}
\label{Fig:stat:ratio-corr}
\end{figure}

Figure~\ref{Fig:stat:ratio-corr} shows the ratio $R^{(q)}(T; q)$ at $T=120$ MeV in the $q$-expectation value.
The behavior in the case of $q$-expectation value is similar to that in the case of the conventional expectation value. 
The effect of the distribution on the correlation in the $q$-expectation value is weaker than that in the conventional expectation value. 

\begin{figure}
\begin{center}
\includegraphics[width=0.45\textwidth]{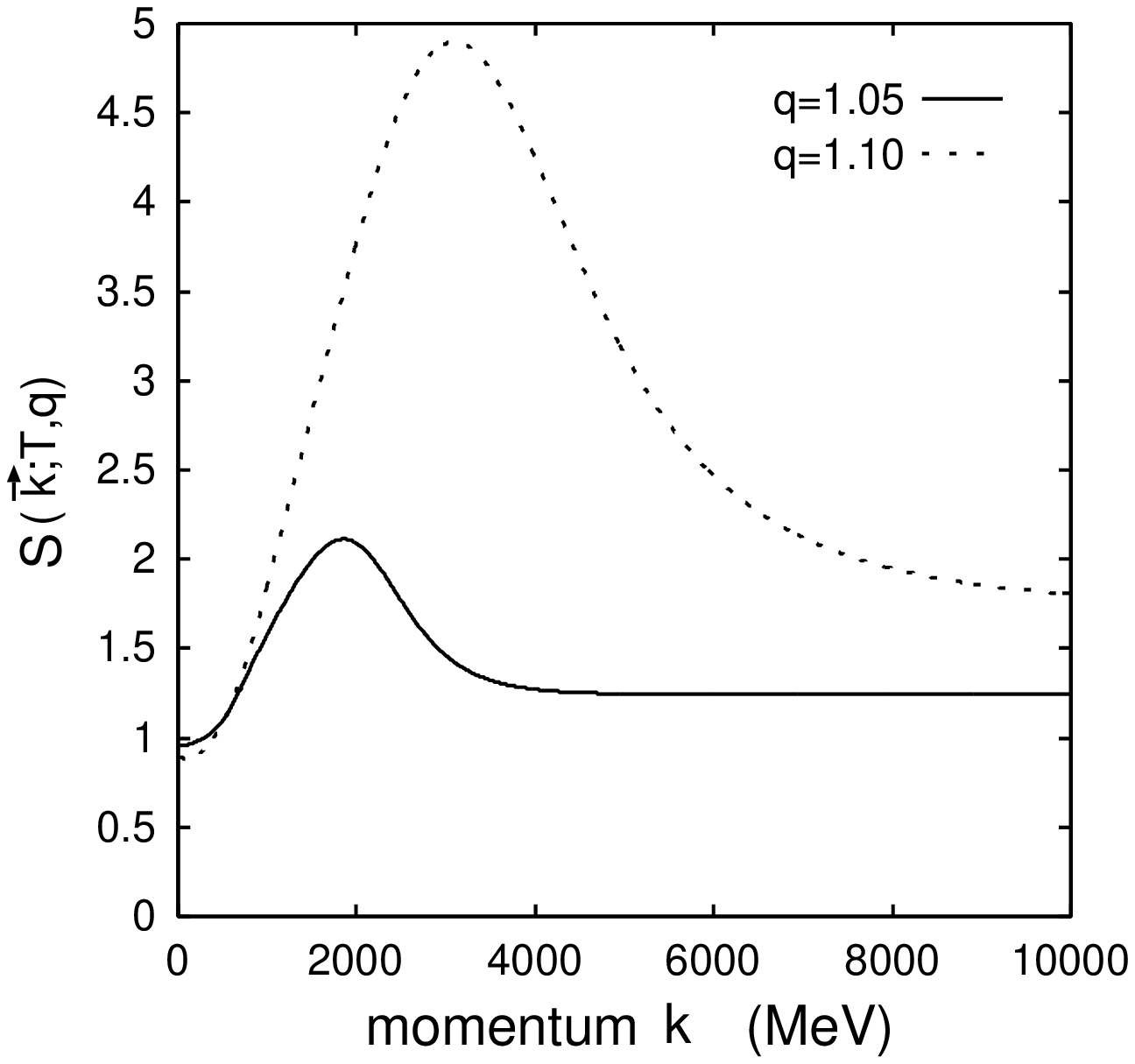}
\includegraphics[width=0.45\textwidth]{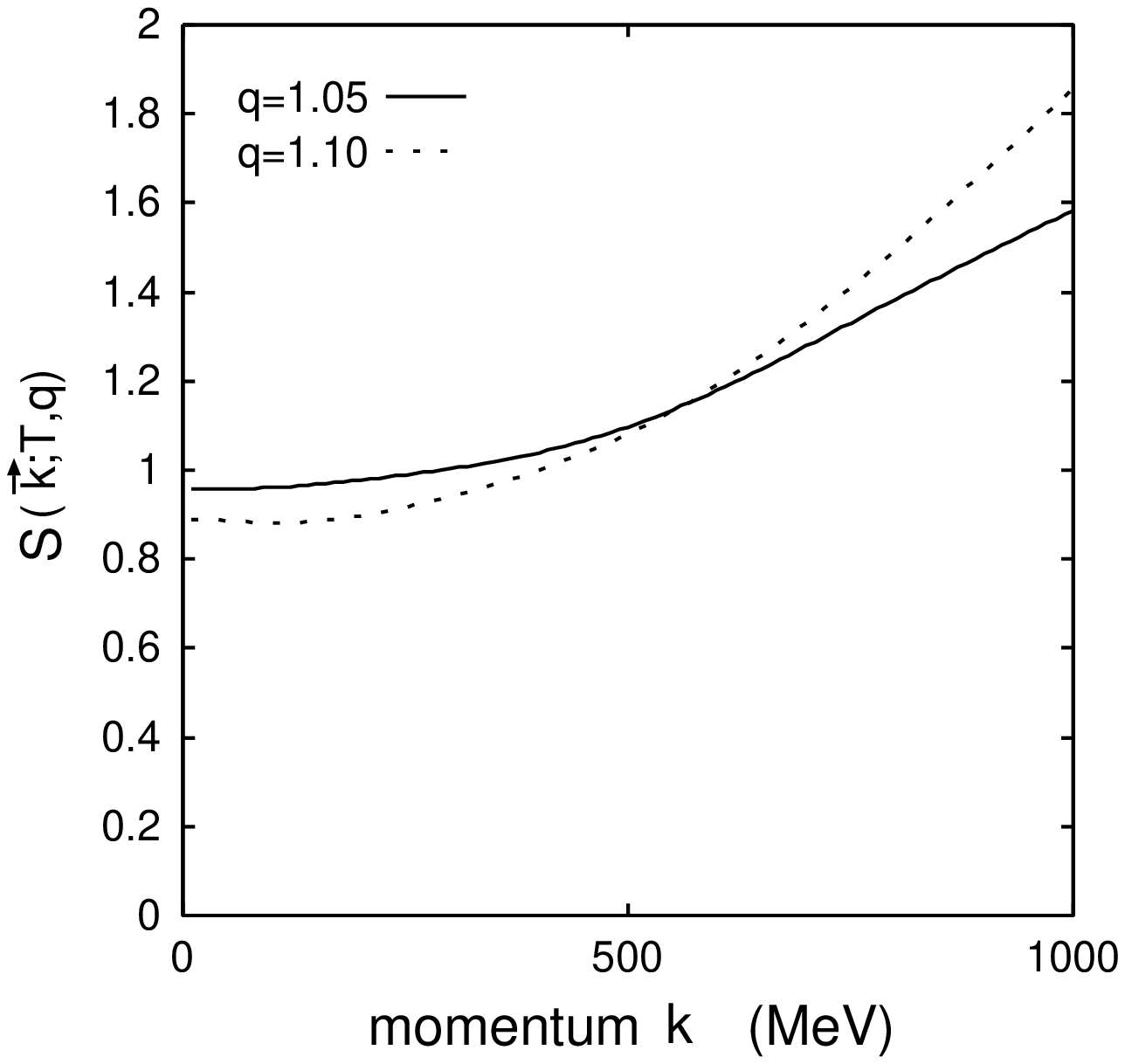}
\end{center}
\caption{
The ratio $S(\vec{k}; T, q)$ at $T=120$ MeV for $q=1.05$ and $1.10$. 
The range of $k$ is $0$ MeV $< k \le 10000$ MeV in the left panel,
and the range of $k$ is $0$ MeV $< k \le 1000$ MeV in the right panel.
}
\label{Fig:ratio-T120-comparison}
\end{figure}

The difference in the correlation between the conventional and $q$-expectation values 
is estimated by the quantity $S(\vec{k}; T, q)$ which is defined by 
\begin{align}
S(\vec{k}; T, q) := \frac{C^{(q)}(\vec{k}, - \vec{k} ; T, q)}{C^{(1)}(\vec{k}, - \vec{k} ; T, q)} . 
\end{align}
The quantity $S(\vec{k}; T, q)$ as a function of $k$ is given in Fig.~\ref{Fig:ratio-T120-comparison}.

The quantity $S(\vec{k}; T, q)$ becomes to be constant as $k$ increases, as shown in Fig.~\ref{Fig:ratio-T120-comparison}, 
This behavior is understood from the the following relation for the pion ($i=1,2,3$):
\begin{equation}
\lim_{|\vec{k}| \rightarrow \infty}  S(\vec{k}; T, q) = 
\frac{\left[ \left(m_{i}^{(1)}(T=0, q)\right)^2  - \left( m_{i}^{(1)}(T, q) \right)^2 \right]^2}{\left[ \left(m_{i}^{(q)}(T=0, q)\right)^2  - \left( m_{i}^{(q)}(T, q) \right)^2 \right]^2 } . 
\end{equation}
We note that $m_{i}^{(1)}(T=0, q)$ equals $m_{i}^{(q)}(T=0, q)$. 
It is apparent from the $T$ dependence of the mass that $S(\vec{k}; T, q>1)$ is greater than one for large $k$, 
as seen in Fig.~\ref{Fig:ratio-T120-comparison}

In the right panel of Fig.~\ref{Fig:ratio-T120-comparison}, 
$S(\vec{k}; T, q)$ is smaller than $1$ when $k$ is sufficiently smaller than $500$ MeV,  
and $S(\vec{k}; T, q)$ is larger than $1$ when $k$ is sufficiently larger than  about $500$ MeV.  
This indicates that the correlation $C^{(1)}$ is smaller than $C^{(q)}$ for small $k$ and that $C^{(q)}$ is larger than $C^{(1)}$ for large $k$.

\section{Discussion and conclusion} 
We investigated the momentum distribution and particle correlation caused by the mass difference,  
when the momentum distribution is given by a Tsallis distribution. 
The conventional expectation value and the $q$-expectation value were applied to calculate the effective masses.

The shape of the momentum distribution in the $q$-expectation value is similar to that in the conventional expectation value,  
and the shape of the correlation in the $q$-expectation value is also similar to that in the conventional expectation value.  
These similarities are explained by the fact that the values of the quantities are determined by the mass.

The $q$-dependence of the momentum distribution and the $q$-dependence of the correlation for soft modes 
are different from those for hard modes, respectively.  
The $q$-dependence of the momentum distribution is quite weak for soft modes.
For hard modes, the magnitude of the momentum distribution increases as $q$ increases. 
The correlation at $q>1$ is larger than that at $q=1$ for soft modes 
which indicates that the momentum is lower than about 500 MeV  in the present case, 
while the correlation at $q>1$ is smaller than that at $q=1$ for hard modes 
which indicates that the momentum is larger than about 500 MeV in the present case. 
The $q$-dependence of the correlation for the soft mode is different from that for the hard mode.

The difference due to the difference of the definition of the expectation value were found:
(1) The momentum distribution in the case of the $q$-expectation value is smaller than 
that in the case of the conventional expectation value, 
and 
(2) the particle correlation in the case of $q$-expectation value is weaker than 
that  in the case of conventional expectation for soft modes, 
while the particle correlation in the $q$-expectation value is stronger than that in the case of conventional expectation for hard modes. 
These results come from the fact that
the $q$-dependence of the mass in the case of $q$-expectation value is weaker than that in the case of conventional expectation value.  

In summary,
the magnitude of the momentum distribution for hard modes increases as $q$ increases, 
while the $q$-dependence of the momentum distribution is quite weak for soft modes.
A particle with momentum $\vec{k}$ correlates with a particle with $-\vec{k}$, 
and the magnitude of the correlation decreases for soft mode and increases for hard mode, as $q$ increases.
The $q$-dependences of these quantities in the case of $q$-expectation value are weaker than those in the case of the conventional expectation value, respectively.

We hope that this work is helpful to understand the effects of power-like distributions. 



\begin{thebibliography}{20}
\bibitem{wilk2007} G.~Wilk, Brazillian Journal of Physics \textbf{37}, 714 (2007) . 
\bibitem{osada2008} T.~Osada and G.~Wilk, Phys.~Rev.~C \textbf{77}, 044903 (2008) .
\bibitem{cleymans2012} J.~Cleymans and D.~Worku, J.~Phys.~G: Nucl.~Phys. \textbf{39}, 025006 (2012) .
\bibitem{marques2015} L.~Marques, J.~Cleymans, and A.~Deppman, Phys.~Rev.~D \textbf{91}, 054025 (2015) .


\bibitem{tsallis-book} C.~Tsallis, {\it Introduction to Nonextensive Statistical Mechanics} (Springer Science+Business Media, LLC, 2010). 

\bibitem{tsallis1998} C.~Tsallis, R.~S.~Mendes, and A.~R.~Plastino, Physica~A \textbf{261}, 534 (1998) .
\bibitem{lavagno2002} A.~Lavagno, Phys.~Lett.~A \textbf{301}, 13 (2002) .

\bibitem{Santos2014} A.~P.~Santos, F.~I.~M.~Pereira, R.~Silva, and J.~S.~Alcaniz, {J.~Phys.~G:~Nucl.~Part.~Phys.} {\bf 41}, 055105 (2014).
\bibitem{Rozynek2009} J.~Ro\.zynek and G.~Wilk, {J.~Phys.~G:~Nucl.~Part.~Phys.} {\bf 36}, 125108  (2009) .

\bibitem{ishihara2016} M.~Ishihara, Int.~J.~Mod.~Phys.~E \textbf{25}, 1650066 (2016) .
\bibitem{ishihara2015} M.~Ishihara, Int.~J.~Mod.~Phys.~E \textbf{24}, 1550085 (2015) .

\bibitem{birrell:book} N.~D.~Birrell and P.~C.~W.~Davies, {\it Quantum fields in curved space} (Cambridge University Press, 1982) 

\bibitem{asakawa1999} M.~Asakawa, T.~Cs\"or\H{o}, and M.~Gyulassy, Phys.~Rev.~Lett. \textbf{83}, 4013 (1999) .
\bibitem{padula2006} Sandra~S.~Padula, G.~Krein, T.~Cs\"or\H{o}, Y.~Hama, and P.~K.~Panda, Phys.~Rev.~C \textbf{73}, 044906 (2006) .

\bibitem{sinyukov1994} Yu.~M.~Sinyukov, Nuclear Physics A \textbf{566}, 589c (1994) .
\bibitem{sinyukov1999} Yu.~M.~Sinyukov, Heavy Ion Physics \textbf{10}, 113 (1999) . 

\bibitem{gavin1994} S.~Gavin and B.~M\"uller, Phys.~Lett.~B \textbf{329}, 486 (1994) .
\bibitem{ishihara1999} M.~Ishihara and F.~Takagi, Phys.~Rev.~C \textbf{61}, 024903 (1999) .

\end{thebibliography}
\end{document}